\documentclass[12pt,preprint]{aastex}
\usepackage{graphicx}
\usepackage{natbib}
\bibpunct[, ]{(}{)}{,}{a}{,}{,}

\def\kms.{\hbox{$\,\rm km\,s^{-1}$}}
\def\ms.{\hbox{$\,\rm m\,s^{-1}$}}

\title{Detection of Large Grains in the Coma of
  Comet~C/2001~A2~(LINEAR) from Arecibo Radar Observations}
\author{Michael C. Nolan, John K. Harmon, Ellen S. Howell\\
Arecibo Observatory, HC3 Box 53995\\
Arecibo PR 00621, USA
\and 
Donald B. Campbell, Jean-Luc Margot\\
Astronomy Department, Cornell University\\
Ithaca NY 14853, USA}

\begin{document}
\section*{}
Proposed Running Head:
Radar Detection of Large Coma Grains

Contact Address:

Michael C. Nolan

Arecibo Observatory

HC 3 Box 53995

Arecibo PR 00612 USA

Phone: (787) 878 2612 x 334

Fax: (787) 878 1861

nolan@naic.edu
\newpage

\section*{Abstract}
  Arecibo S-band ($\lambda=13\,$cm) radar observations of Comet C/2001
  A2 (LINEAR) on 2001 July 7--9 showed a strong echo from large coma
  grains. This echo was significantly depolarized. This
  is the first firm detection of depolarization in a grain-coma radar echo
  and indicates that the largest grains are at least $\lambda / 2$ or
  2 cm in radius. The grains are moving at tens of $\rm m\,s^{-1}$ with respect to the nucleus. The non-detection of
  the nucleus places an upper limit of 3 km on its diameter. The
  broad, asymmetric echo power spectrum suggests a fan of grains that have a
  steep (differential number $\sim a^{-4}$) size distribution at cm-scales, though the observed fragmentation of this comet complicates that picture.

Keywords: radar, comets
%\end{abstract}
\newpage
\section{Introduction}
When comets enter the inner solar system, we get a close look at material that is relatively pristine, closely resembling the volatile-rich material that formed the outer planets and their satellites. Comets may also have contributed significant volatiles to the early terrestrial planets during accretion, although the relative contribution of cometary impactors and more rocky asteroidal bodies is still debated. Comets also present an impact hazard for the Earth.
Understanding the physical structure of cometary nuclei and how they
fragment is important for learning about their formation and
evaluating the impact hazard. The break-up of Comet D/1993 F2
(Shoemaker-Levy 9) has reinforced the idea that comets may be
aggregates of icy grains that easily fragment into similar-sized
pieces (e.g., Asphaug and Benz, 1996\nocite{AsphaugBenzSL9}).

The existence of large icy grains in cometary comae has long been inferred. Solar radiation pressure affects the orbital motion of small particles \citep{Burns79}. By examining the orbital evolution of grains, it is possible to determine their size, based on the extent to which solar radiation affects their orbits. The observed motions of particles in comet tails are consistent with sizes of mm to cm in a number of cases. Modeling of comet tails suggests large particles in some cases (summarized in Fulle, 2004). The orbital motion of some of dust bands also require that the particles be mm to cm in size (e.g., Sykes et al., 1986, Reach et al., 2000)\nocite{Reach00:80}.

Submillimeter observations of C/1996 B2 (Hyakutake) showed evidence of
emission from mm-scale particles \citep{Jewitt97:1145}. The
distributions of CO and formaldehyde are often found to be
inconsistent with a purely nuclear source, and sublimation of grains
is inferred (e.g., Eberhardt, 1999, Gunnarson, 2003). However, it is not known whether these grains are an important source of other gas species or dust. Radar observations have indicated the presence of large grains in comet Halley (Campbell et al., 1983) and Iras-Araki-Alcock (Harmon et al., 1983), though the grains were not directly detected. Radar observation of large (cm-sized) grains near the nucleus determines their velocity distribution, improving our understanding of how grains leave the nucleus, which in turn helps determine the relative contributions of the nucleus and the grains to the gas coma. 

This paper will give a brief description of earlier radar results (Section 2), present our observations of C/2001~A2, and use those observations to constrain the size of the nucleus and the population of coma grains (Section 3). We then interpret the grain population both in the context of a simple gas-drag ejection model, yielding a size distribution of the coma grains that contribute to the radar echo (Section 4), and as the outcome of earlier fragmentation of the nucleus (Section 5). We compare the results to those from other comets and using other methods (Section 6). We then conclude (Section 7) that the grain size distribution determined in the modeling is likely to be correct even if the model is oversimplified.

\section{Background}
If a comet nucleus comes close enough to the Earth, we may have the opportunity to obtain 2-dimensional radar delay-Doppler images, which can provide a wealth of information about the comet's shape, size, and surface characteristics. However, because the signal-to-noise ratio (SNR) of radar observations is very sensitive to the distance $\Delta$ to the target (${\rm SNR}\sim \Delta^{-4}$), radar imaging of comets requires an approach distance of 0.1 AU or less. Over the last twenty years, only comets C/1983~H1 (IRAS-Araki-Alcock) and C/1983~J1 (Sugano-Saigusa-Fujikawa) in 1983 and comet C/1996~B2 (Hyakutake) in 1996 have approached within this distance from the Earth, but radar imaging was not technically feasible for any of those objects.

Even if imaging is not possible, we can measure the basic cm-wavelength scattering properties and estimate the size of the nucleus of any comet that comes within about 0.5 AU of the Earth. Altogether, nine comets have been observed with radar, all at Arecibo except for IRAS-Araki-Alcock, which was observed at both Arecibo and Goldstone, and Hyakutake, which was observed only at Goldstone. All of these objects were observed by transmitting a monochromatic signal and measuring the Doppler shift and broadening of the return echoes. This broadening can yield rotational information for the nucleus and particle velocities for coma grains. We measure both circular polarizations of the returned echo, because the polarization properties of the echo can yield surface roughness information for the nucleus and information on coma grain sizes. Of the nine comets successfully observed, the nucleus was detected in six (all but Halley, C/2001~A2, and C/2002 O6) and coma grains were detected in five \citep{Harmon04}.

\begin{figure}
\begin{center}\includegraphics[width=5in]{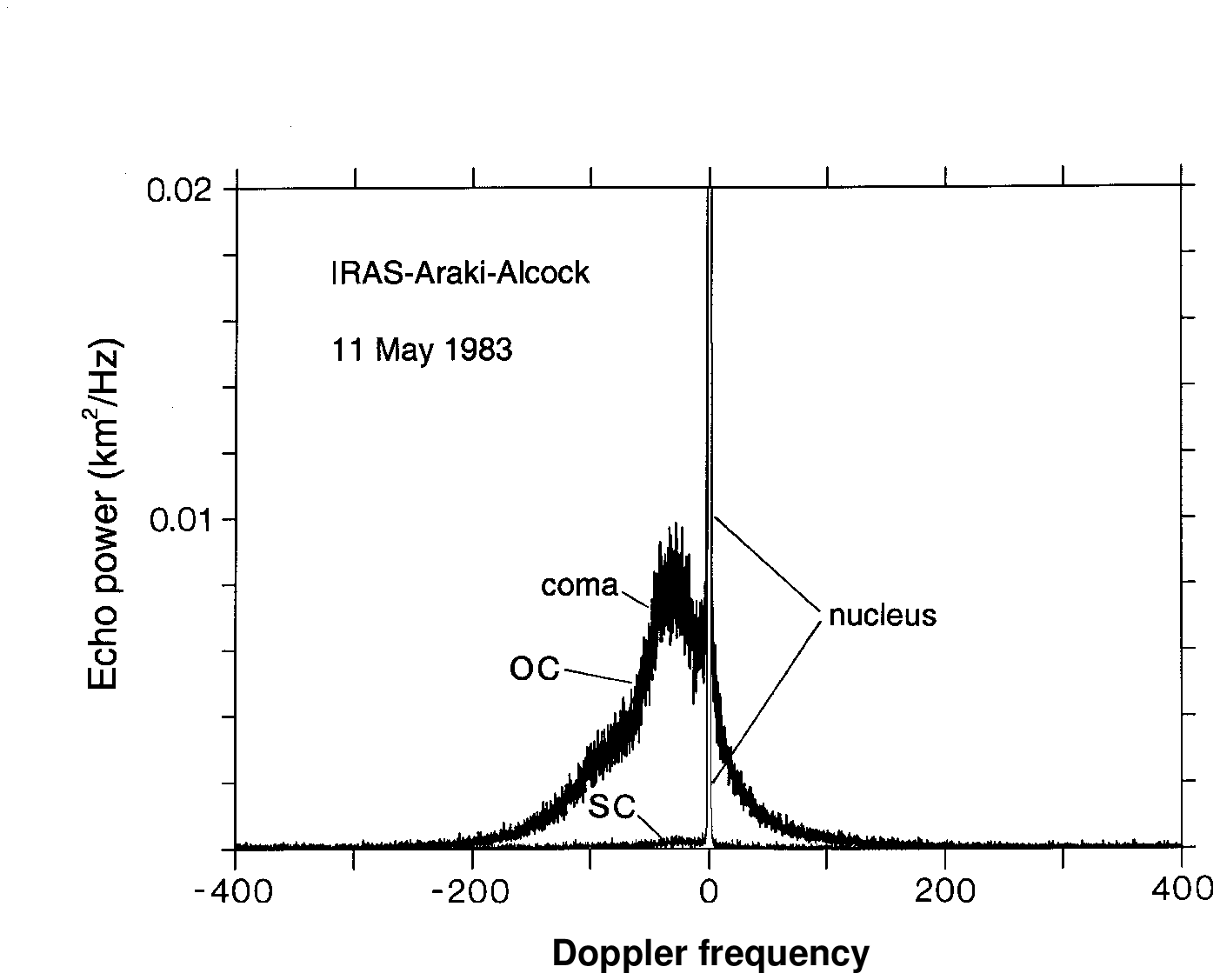}\end{center}
  \caption{\label{fig:IAA} Doppler spectra
    (OC and SC polarizations) for Comet C/1983 H1 (IRAS-Araki-Alcock)
    show the narrow nucleus echo and broad grain-coma echo. The
    SC/OC polarization ratio of the coma is estimated to be $\mu_{\rm
      c} = 1.4$\%,
    after correcting for instrumental crosstalk. The spectrum has been
    truncated so that only the bottom 2\% of the nucleus echo
    show. These data are from Arecibo S-band ($\lambda = 13\,$cm)
    radar observations on 1983 May 11 \citep{Harmon89:1071}.}
\end{figure}

For a cometary echo, two different mechanisms introduce Doppler broadening. These are illustrated in Fig.~\ref{fig:IAA}, which shows the echo power spectrum of IRAS-Araki-Alcock taken at Arecibo in 1983 (Harmon et al., 1989).
The central spike is an echo from the solid comet nucleus. Its frequency width $\delta\nu$ is proportional to the rotation rate and diameter of the nucleus:
\begin{equation}
\label{eq:rot}
\delta\nu = 2 {\delta v \over c} \nu = {4 \pi D \nu \sin i \over P c} = {4 \pi D \sin i \over P \lambda},
\end{equation}
where $\delta v$ is the velocity difference, with respect to the observer, between the approaching and receding edges, $P$ is the rotation period, $D$ is the nucleus equatorial diameter in the plane of the sky, $i$ is the angle between the comet's spin axis and the line of sight, $\nu$ is the observing frequency, $\lambda$ is the observing wavelength, and $c$ is the speed of light. The broader echo component from grains in the coma and its frequency width is due to the velocity dispersion (projected along the line of sight) of all coma particles within the beam that contribute to this part of the radar echo.

For IRAS-Araki-Alcock, the reflected power from the coma was almost entirely ($98.6$\%) in the opposite circular polarization (OC) from that transmitted (see Fig.~\ref{fig:IAA}), suggesting single scattering from small (Rayleigh) scatterers no larger than about 2 cm in size, but not much smaller, either \citep{Harmon89:1071}. The echo from the nucleus still shows most (90\%) of the echo in the OC, with the other 10\% in the same sense of circular polarization (SC) as that transmitted.  Multiple scattering or reflection from a surface that is rough at wavelength scales has ``depolarized'' a portion of the echo power. The IRAS-Araki-Alcock observations were made when the comet was very close (0.033 AU) to the Earth, resulting in extremely strong echoes. No comparable opportunity has presented itself since.
 
\section{Observations of C/2001~A2 (LINEAR)}
Comet C/2001~A2 (LINEAR) was discovered on 2001 January 3 by the LINEAR
near-Earth asteroid discovery program \citep{Stokes00:21}. We
observed the comet on 2001 July 7--10 with the Arecibo S-band radar.
The observing circumstances are summarized in
Table~\ref{tab:radar}.

We transmitted a left-circularly-polarized, monochromatic 2380-MHz radio wave at a power of 900 kW, and observed the echo power spectrum in both left (SC) and right (OC) circular polarizations. We report the radar cross section $\sigma$, defined as the cross sectional area $\pi R^2$ of a metal sphere that would reflect the same power as we measure. The radar albedo is then the ratio of the radar cross section to the physical cross section of the target. The polarization ratio $\mu_{\rm c}$ is the ratio $\sigma_{\rm sc} / \sigma_{\rm oc}$.

At the time of observations, the Arecibo telescope had a forward gain of 73 dB. The receiving system had an effective collecting area of 9 K/Jy (25000 m$^2$) and a system noise temperature of about 25K. The gain, effective area, and system temperature vary by about 20\% with telescope pointing, and we used values based on the actual pointing in data reduction. Every 10 s, we changed the transmitter frequency by 10 kHz, with a corresponding change applied in the data processing, in order to obtain a measurement of the background so that it could be accurately subtracted.

It is not feasible to simultaneously transmit and receive with the same antenna, so we transmitted for the round-trip light time to the comet and then, shortly before the reflected light arrived back at the antenna, turned off the transmitter and received the returned echo for the round-trip light time to the comet. The radial component of the overall (ephemeris) velocity of the comet was removed in the datataking process by introducing small offsets in the frequency of the transmitted signal to compensate for the Doppler shift of the center of mass of the target as predicted by an ephemeris based on previous optical observations.

C/2001~A2 fragmented several times before (and perhaps after) our observations. For these observations, the pointing and Doppler ephemerides were for the ``B'' fragment, at that time apparently the only visible fragment (Sekanina et al., 2002), though other fragments and their remnants may have been in the beam, particularly fragments D, E, and F (see Section 5).

\begin{figure}
% 07: 70 429 526 693
% 08: 174 434 526 693
% 09: 70 377 526 651
% sum: 151 377 526 646
\hbox{\hbox to 30 pt{\vbox{\vfill
\includegraphics[scale=.5,bb=70 256 100 804]{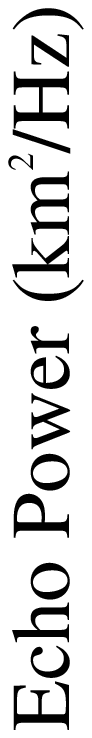}\vfill}}\hbox{\vbox{\hbox {\includegraphics*[scale=.5,bb = 100 429 526 693]{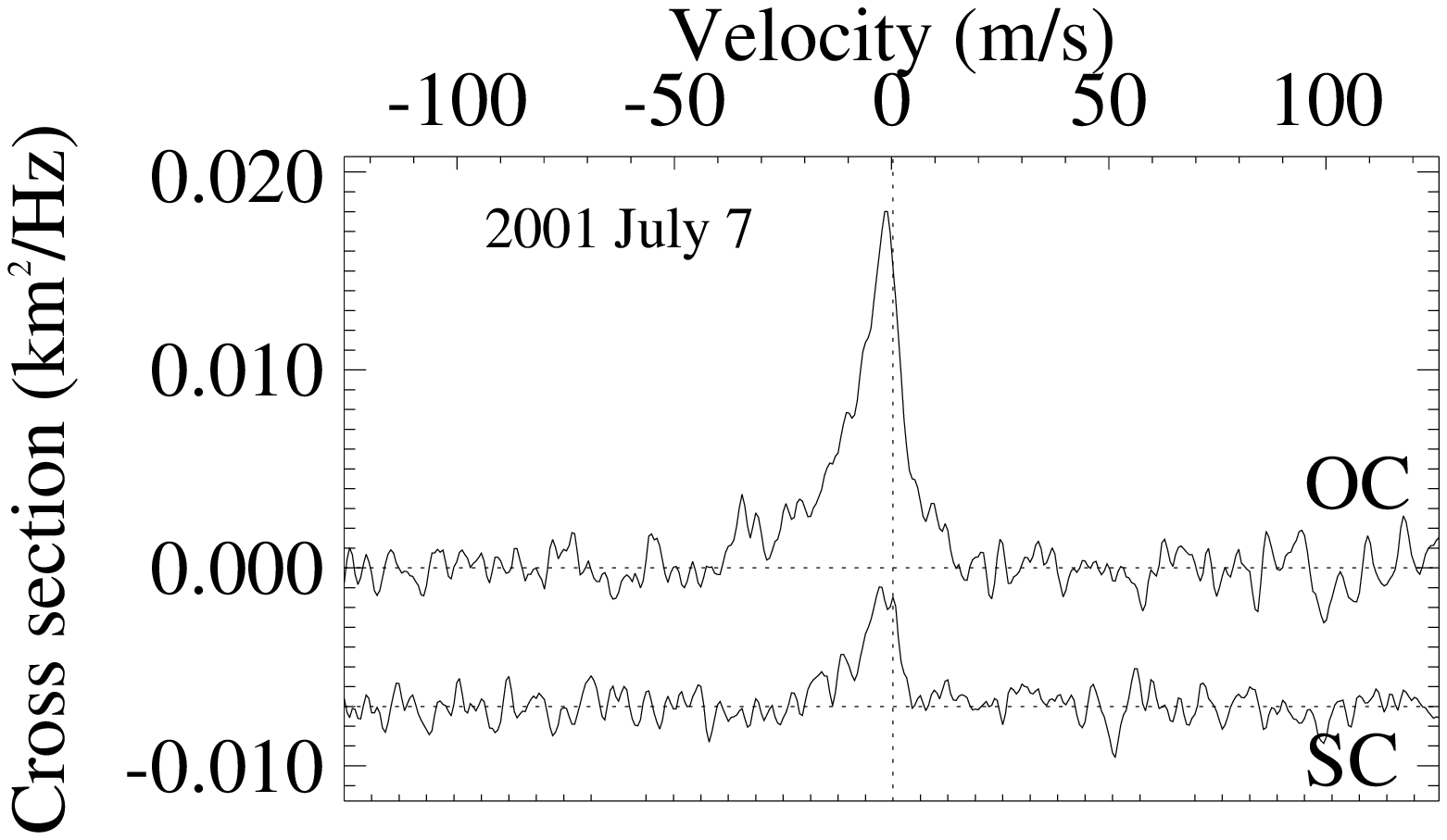}
%\hskip -1 cm 
\includegraphics[scale=0.5,bb = 174 429 526 693]{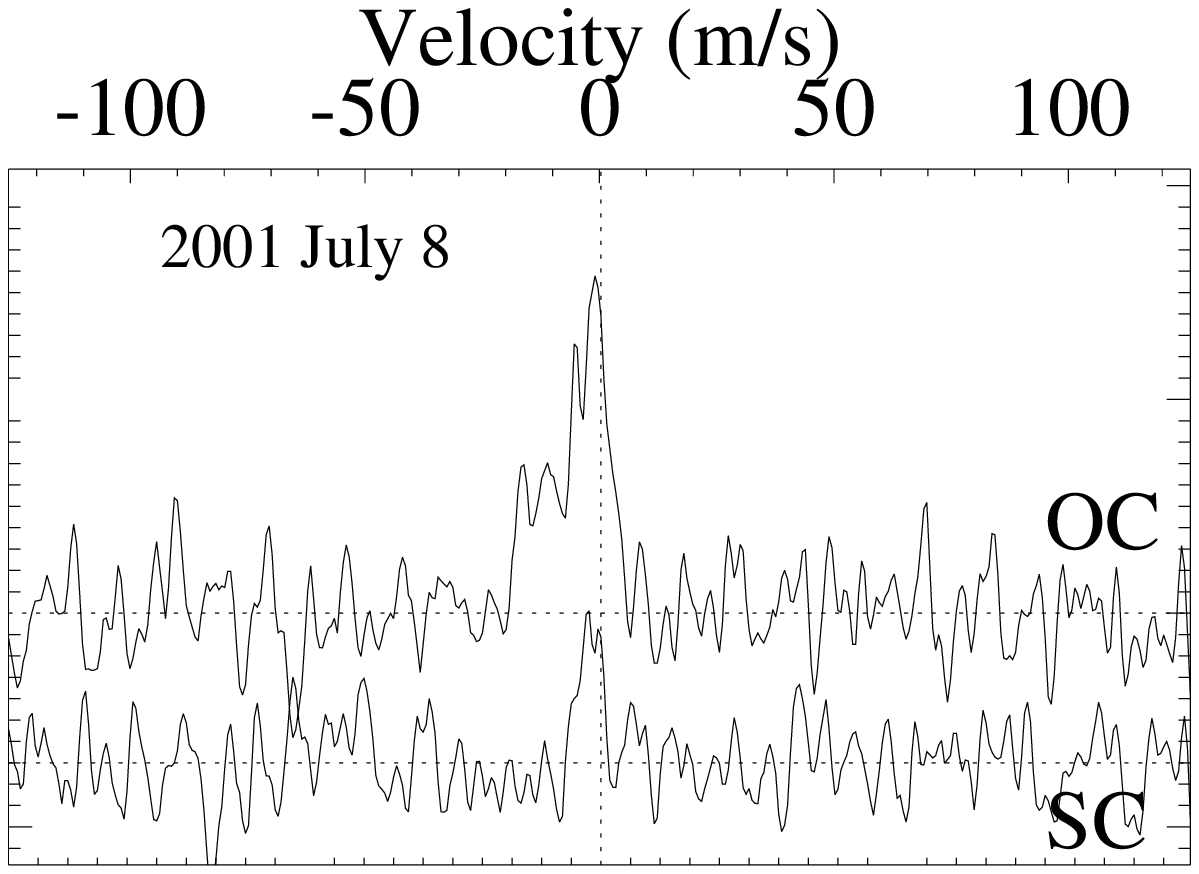}\hfill}
%\vskip -2 cm
\hbox to \hsize{\includegraphics*[scale=.5,bb = 100 377 526
  651]{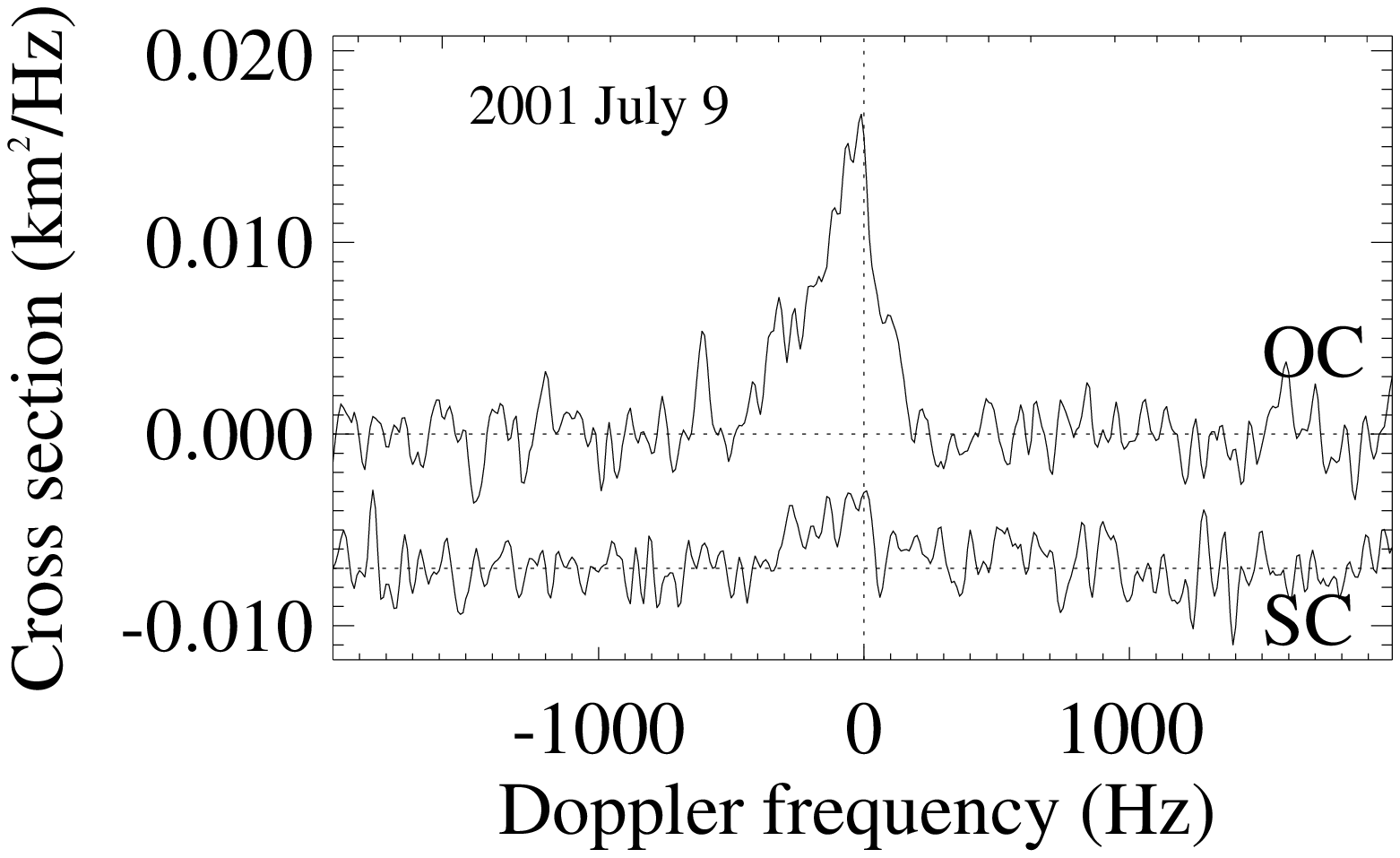}
%\hskip -1 cm
\includegraphics[scale=.5,bb = 174 377 526 651]{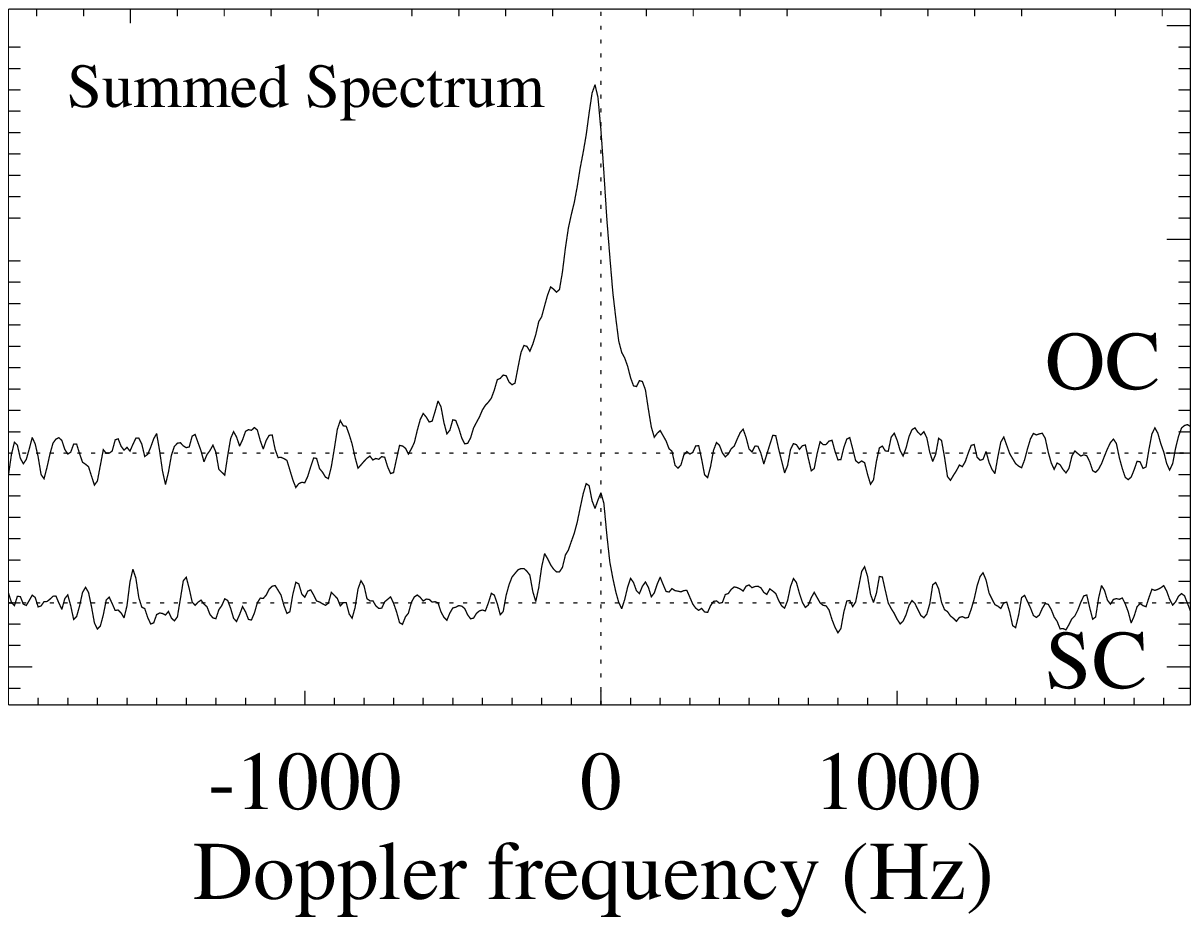}\hfill}}}}
\caption{\label{fig:A2spec}Echo power spectrum of C/2001~A2 on 2001
  July 7--July 9, and the variance-weighted sum of all three days.
  The top line shows the OC and the bottom line the SC polarization.
  The SC spectra have been vertically offset by -0.007 km$^2$/Hz for clarity.  The spectra were taken at a resolution of 10 Hz, and then smoothed to 40 Hz.}
\end{figure}

Echo power spectra of C/2001~A2 on 2001 July 7--9 are shown in Fig. \ref{fig:A2spec}, and their parameters are summarized in Table \ref{tab:radar}. The relatively broad OC echo power spectrum peaked near the predicted ephemeris Doppler shift, and had a width of at least 900 Hz (57 \ms.) on each date. The echo is asymmetric, with more power in the receding (negative Doppler shift) direction. Asymmetry was also clear in the coma echoes from IRAS-Araki-Alcock (Fig.~\ref{fig:IAA}) and Hyakutake (Harmon et al., 1997).

The SC echo power is about 25\% that of the OC power. Its spectrum appears to have a somewhat narrower bandwidth than that of the OC echo, although this apparent difference may be due to the lower SNR. 

\subsection{Measurement Uncertainty}
The uncertainties in the reported values are dominated by calibration errors, not random noise: The thermal noise is $< 5$\% for $\sigma_{\rm oc}$ on each day. The systematic errors in the cross sections include calibration uncertainties for the transmitter power, the antenna performance, and the receiver thermal noise, and total to about 30\%. In the polarization ratio $\mu_{\rm c}$, all but the receiver noise calibration uncertainties cancel, and the systematic uncertainty is only about 5\%. The day-to-day variations in these systematic calibration errors are also smaller, about 5--10\%. The day-to-day variation in the OC and total cross section is $20-30$\%, and thus statistically significant, but the other parameters are not.

\subsection{Nucleus or Coma?}
Our conclusion depends on the observed SC echo being reflection from
coma grains, so we first examine the hypothesis that the echo could
come from the nucleus instead. These spectra do not show the ``spike
plus skirt'' pattern that distinguished the coma from the nucleus echo
in (for example) the IRAS-Araki-Alcock spectra. The SC echo appears to
be at least 300 Hz wide. The rotation period of the nucleus has been
estimated at either 6 or 3 hours by \citet{Woodney01:1121}. Assuming
the same radar albedo as IRAS-Araki-Alcock (4\%, the lowest
well-determined comet radar albedo \citep{Harmon99}), a nucleus larger
than 3 km diameter with rotation period of at least three hours would
have been detectable with a few Hz bandwidth at $>10 \sigma$ on July
7, The strongest narrow features that could be from the nucleus are
present only at the 3-$\sigma$ significance level, so we consider 3 km
to be an upper limit to the nucleus diameter. In order to have a
bandwidth of 300 Hz, the nucleus would have to be at least 30 km in
diameter, which would have been visible optically about 5 magnitudes
brighter than was actually observed in January and February 2001.
Thus, from the large bandwidth and overall echo spectrum shape we
conclude that the echo power was dominated by reflections from coma
grains. The nucleus may contribute to the peak of the echo, but it
cannot be distinguished.

\subsection{Coma Echo}
The spectra in Fig.~\ref{fig:A2spec} show that there is a substantial
SC echo. Although this degree of depolarization is typical of comet
nuclei, we have shown that the SC echoes cannot have come from the
nucleus. Hence, we conclude that significant depolarization ($\mu_{\rm
  c} = 28\%$) was introduced in the radar backscatter from large-grain
coma. This is in stark contrast to the coma echo from
IRAS-Araki-Alcock (Fig.~\ref{fig:IAA}), which had $\mu_{\rm c}$ only
1.4\%. Cometary grain comae are too sparsely populated to give
significant depolarization from multiple scattering
\citep{Harmon89:1071,Harmon04}, so any detectable depolarization must
arise from the single scattering off individual irregular grains.
Depolarization from irregular grains is small ($\sim$ 0.1--3\%) for
sizes in the Rayleigh ($a < \lambda/2\pi$) regime and increases
rapidly to large values ($\sim$ 10--50\%) as the radius increases
beyond the $\lambda/2\pi$ threshold \nocite{Harmon89:1071}(Harmon et
al., 1989). The precise depolarization behavior depends on the grain
density and degree of irregularity. For example, the transition size
can be somewhat larger than $\lambda/2\pi$ for low-density grains. If
the differential grain size distribution is described by a power law
$N_a \propto a^{-\alpha}$ with a cutoff radius $a_{\rm max}$, then the
strong Rayleigh dependence for the grain backscatter efficiency will
tend to make the echo depolarization be dominated by the larger grains
unless the size distribution is incredibly steep ($\alpha > 5$), so
that $\mu_{\rm c}$ can be quite sensitive to whether $a_{\rm max}$ is
larger or smaller than $\lambda/2\pi$. For IRAS-Araki-Alcock, the case
was made, based on total-mass arguments and the low depolarization,
that $a_{\rm max}$ must have been of the order of, but not much
smaller than, $\lambda/2\pi$, which is 2 cm for the 13 cm wavelength
used \nocite{Harmon89:1071}(Harmon et al., 1989), and that a cutoff
size of this order was consistent with the gravity-limit against
gas-drag (see Section~\ref{sec:model}). A comet with a smaller nucleus
size or higher gas flux than IRAS-Araki-Alcock could have $a_{\rm
  max}$ larger than $\lambda/2\pi$, and hence show significant
depolarization. Both Halley \nocite{Campbell89:1094}(Campbell et al.,
1989) and Hyakutake \nocite{Harmon97:1921}(Harmon et al., 1997) did,
in fact, show hints of non-negligible depolarization, although in both
cases the SNR was too low to be certain. Thus, the new depolarization
results for C/2001~A2 provide the first unambiguous radar evidence for
large grains with sizes of several centimeters or more.

If the SC spectrum really is narrower than the OC spectrum, it may be
due to a dispersion in ejection velocity as a function of grain size,
because only large particles ($>$ 2 cm) give substantial SC echo, and
larger particles are likely slower moving (see
Section~\ref{sec:model}). Note that, unlike the case of a rotating
solid body, there is no direct spatial information in the Doppler
spectrum of the coma: The coma is assumed to fill the radar beam, and
we observe the echo of the entire population of grains.

The radar cross-section shows day-to-day variations at the 20\% level.
These variations probably represent real differences in the grain
population. There is also a hint of a ``shoulder'' on the
low-frequency tail of the spectrum. Since small grains are more easily
accelerated to higher velocities than large ones, it is possible that
this shoulder appears where the grain size approaches the Rayleigh
scattering limit (2 cm for these observations), reducing the
scattering efficiency, and thus the cross section at high velocity.
The spectrum of C/2002~O6 (SWAN) shows a similar shoulder
\citep{Harmon04}, but in both cases, the data are too noisy to address
this idea in detail.

\section{Grain-Coma Modeling\label{sec:model}}

\subsection {Grain Ejection}
We have modeled the grain coma in order to estimate the characteristic grain velocity, place constraints on the grain ejection direction, and estimate the nucleus mass-loss rate in grains. We adopt the approach taken in modeling the coma radar echoes from comets IRAS-Araki-Alcock (Harmon et al., 1989) and Hyakutake (Harmon et al., 1997). The starting point is the canonical gas-drag model for dust release originally proposed by Whipple (1951) and adapted later by others (see Harmon et al., 1989, Harmon et al., 2004, and references therein). 

According to this theory, the radial gas drag from an outgassing nucleus of radius $R$ accelerates a grain of radius $a$ from the surface to a terminal velocity $V_{\rm t}$ given by
\begin{equation}
\label{eq:Vterm}
   V_{\rm t}(a) = C_{\rm v}\, a^{-1/2}\, (1-a/a_{\rm m})^{1/2}\;\;\;\;,
\end{equation}
where
\begin{equation}
\label{eq:Cv}
   C_{\rm v} = \left(\frac{3 C_{\rm D} V_{\!\rm g} Z R}
               {4\rho_{\rm g}}\right)^{\!\! 1/2}
\end{equation}
is a velocity scale factor and
\begin{equation}
\label{eq:am}
   a_{\rm m} = \frac{9 C_{\rm D} V_{\!\rm g} Z}
               {32\pi G R\, \rho_{\rm n} \rho_{\rm g}} =
               \frac{3 C_{\rm v}^2}{8\pi G R^2 \rho_{\rm n}}
\end{equation}
is the radius of the largest grain that can be lifted off the surface against nucleus gravity.  Here, $C_{\rm D}$ is a drag coefficient, $V_{\!\rm g} = 1.7\times 10^3\, T({\rm Kelvin})^{1/2}\rm\,cm\,s^{-1}$ is the gas velocity, assumed to be the thermal expansion velocity at temperature $T$, $Z$ is the gas mass flux at the surface, $G$ is the gravitational constant, $\rho_{\rm n}$ is the mean density of the nucleus, and $\rho_{\rm g}$ is the density of a grain.

Since grains are presumably accelerated in jets, where the gas velocity is considerably higher than the simple thermal expansion velocity, the scale factor $C_{\rm v}$ will tend to be larger for comets with more violent jetting. 

We assume simple forms for the grain ejection directions and grain
size distributions. The grains are ejected in a conical fan, and the
grain radii are assumed to be power-law distributed as
$a^{-\alpha}$. Again, the grains are presumably ejected from
illuminated jets, not a uniform cone. We compute the Doppler
spectrum from the radial components of the terminal velocities of the
entire ensemble of grains within the radar beam, assuming a stable fan
of continuously ejected grains, and accounting for their orbital
motion. We then adjust $C_{\rm v}$ and fan direction to match the
width and asymmetry of the observed spectrum.

In broad terms, the size distribution of the particles determines the
shape of the spectrum: a steep distribution will have many small
grains, which can be accelerated to a high velocity, and give a broad
``skirt'' to the spectrum. The velocity scale factor $C_{\rm v}$
determines the actual width of the spectrum, as long as the size
distribution is steep enough that the small particles dominate the
echo. A fan (as opposed to isotropic) shape is required to give an
asymmetric spectrum. Changing the cone angle of the fan affects the
spectrum width, with a narrower cone angle tending to reduce the
breadth of the spectrum. However, we find that we need a broad cone
angle (120$^\circ$), a steep size distribution and a large $C_{\rm v}$
to model the rather broad observed spectrum, so that the effects of
adjusting the cone angle will be minor, and cannot be easily
distinguished from those of adjusting the jet orientation.

The numerical results that follow assume that the grain density
$\rho_{\rm g}=0.5\, \rm g cm^{-3}$ and the nucleus density $\rho_{\rm
n}=1\,\rm g\,cm^{-3}$, surface temperature $T=250\,\rm K$ and radius
$R=1\,$km.

\begin{figure}
\begin{center}
\includegraphics[width=5in]{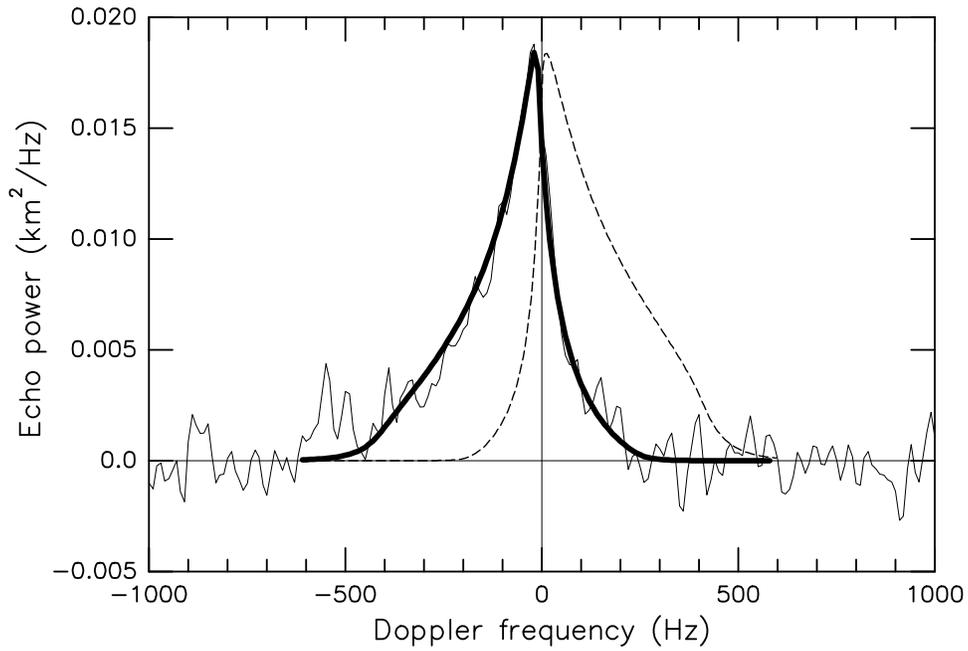}
\end{center}
\caption{\label{fig:model}Radar Doppler spectrum with an overplotted model spectrum (heavy solid curve) for a 120$^{\circ}$ grain ejection fan centered on the near-antisolar direction (165$^{\circ}$, 0$^{\circ}$) and assuming $C_{\rm v} = 36\,{\rm
    cm}^{1/2}$ \ms. and $\alpha = 3.9$. A model spectrum for a sunward
  (0$^{\circ}$, 0$^{\circ}$) fan using these same parameters (short
  dashes) gives a much poorer fit.}
\end{figure}

\begin{figure}
\begin{center}\includegraphics[width=3in]{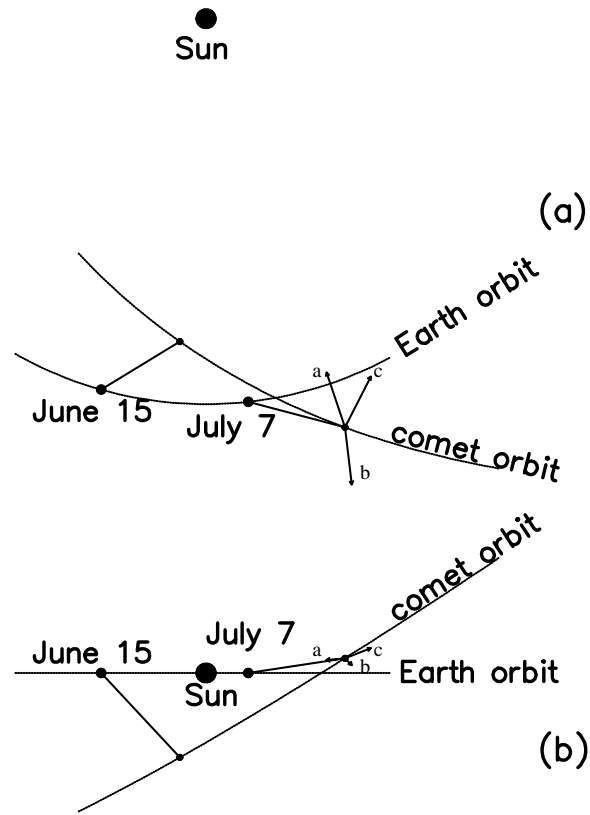}\end{center}
\caption{\label{fig:geom}Orbital geometry for C/2001 A2 (LINEAR)
  during June--July, 2001, as seen (a) looking down on the ecliptic
  plane and (b) in the ecliptic plane. The Earth-comet positions are
  shown for June 15 and at the time of the radar observation on July
  7. Also shown are July 7 cometocentric vectors showing the sunward
  direction (0$^{\circ}$, 0$^{\circ}$) (a), the
  near-antisolar direction (165$^{\circ}$, 0$^{\circ}$) (b),
  and the ($-45^\circ$, $-15^\circ$) direction (c). Both b and c are consistent with the radar data, because they have the same Earth-comet-fan angle of about 100$^\circ$, but a is not.} 
\end{figure}

Assuming a conical ejection fan of width 120$^\circ$ and a power-law
index $\alpha = 3.9$, we get good fits for certain fan directions by
using $C_{\rm v} = 36\, \rm cm^{1/2}\,m\,s^{-1}$. An example of a
model spectrum giving a good fit is shown in Fig.~\ref{fig:model}.
This example corresponds to an Earth-comet-fan direction of
(165$^\circ$, 0$^\circ$). The coordinate pair represents cometocentric
azimuth and elevation relative to the comet orbit plane, with azimuth
given relative to the Sun direction, so that an antisolar tail would
be in the (180$^\circ$, 0$^\circ$) direction. The most sunward fan
direction giving a good fit is ($-45^\circ$, $-15^\circ$).  A
precisely sunward (0$^\circ$, 0$^\circ$) fan does not fit the data
because of its positive Doppler offset, as shown in
Fig.~\ref{fig:model}. The relationships between the Sun, the comet fan
direction and the viewing geometry are illustrated in
Fig.~\ref{fig:geom}, which shows the geometry for the three fan
directions discussed.

While we have discussed only two examples, there is a locus of fan
directions that give a good fit corresponding to Sun-comet-fan angles
varying from 50 to 165$^\circ$, depending on the plane-of-sky position
angle. The radar observes the radial component of velocity of the
grains with respect to the Earth. This radial component depends on the
plane-of-sky direction because of the Keplerian motion of the grain
particles. However, good fits are obtained for all directions with the
same Earth-comet-jet angle, 100$^\circ$ from the Earth-comet direction
(as shown in Fig.~\ref{fig:geom}).

The velocity scale factor $C_{\rm v} = \rm 36\, cm^{1/2}$ \ms.
corresponds to $V_{\rm t} = 36$ \ms. for an $a = 1$ cm grain
(neglecting gravity). Changing the size distribution to an $\alpha =
3.5$ power law reduces the spectrum width to half that for the steeper
power law because it weights the larger, slower grains more heavily.
For this shallower power law, simply increasing $C_{\rm v}$ is
ineffectual at widening the spectrum, as $a_{\rm m}$ increases at the
same time. Hence, a simple gas-drag model suggests a steep ($\alpha
\sim 4$) size distribution near the cm size scale in order to match
the observations. Although the overall dust size spectral index
$\alpha$ averaged about 3.5 or 3.6 for Hyakutake and Halley, it can be
closer to 4 at the largest grain sizes or when the grains are
undergoing fragmentation (Fulle et al., 1995, Fulle et al., 1997).

The 36 cm$^{1/2}$ \ms. scaling factor (for $\alpha = 3.9$) is similar
to that estimated for the coma echo from Comet Hyakutake and
significantly larger than that estimated for IRAS-Araki-Alcock and
Halley, as shown in Table~\ref{tab:comets}. This high $C_{\rm v}$
implies that the gas drag effect for C/2001 A2 is strong and capable
of lifting off large grains of the size needed to account for the
observed depolarization. For example, if one assumes $R = 1$ km and
$\rho_{\rm n} = 1\,\rm g\,cm^{-3}$, then using this $C_{\rm v}$ value
in Eq.~4 gives $a_{\rm m} = 10$ m, which is more than enough to
accommodate a significant population of large, depolarizing
grains. One can see why IRAS-Araki-Alcock would have a much smaller
$a_{\rm m}$ comparable with the Rayleigh transition size
$\lambda/2\pi$, since the smaller $C_{\rm v}$ and larger radius for
that comet ($\sim 10$ km) would make the $C_{\rm v}^2/R^2$ factor in
Eq.~4 about a factor of 100 smaller than for C/2001 A2.  Radio
observations of OH molecules give line widths suggesting that
C/2001~A2 has a comparable gas outflow velocity (about 600 \ms.) to
Halley and Hyakutake \citep{Lovell05,Crovisier02}. Thus, nucleus size
may be the most important contributor to large-grain ejection.

If one inserts $C_{\rm v} = 36\,{\rm cm}^{1/2}$ \ms. into Eq.~3 and
assumes $C_{\rm D} =2$, $\rho_{\rm g} = 0.5\,\rm g\,cm^{-3}$, and a
grain temperature of 250K (implying a gas velocity of 270 \ms.), then
solving for $Z$ gives a value of $7\times 10^{-4}\,\rm
g\,cm^{-2}\,s^{-1}$ for the gas flux, which is similar to that
obtained for Hyakutake (Harmon et al., 1997) and an order of magnitude
larger than the nominal sublimation rate for clean water ice at 1
AU. Such high apparent gas fluxes might be possible if the grain
ejection were explosive or jet-like, also consistent with the observed
OH line width velocity of 600 \ms. \citet{Lovell05}. A high $Z$ would
also give a more reasonable surface active fraction than the extremely
high values obtained when one combines the nucleus size upper limit
and measured gas production rates for this comet with the canonical
$Z$ value for sublimating ice. On the other hand, the observed
velocities could be achieved with a smaller $Z$ if the grains were
extremely fluffy (with a higher drag coefficient and lower grain
density). Also, the gas flux at the nucleus would be irrelevant if
rocket forces from their own outgassing accelerated the grains (as
discussed later).

\subsection{Mass Loss}
\begin{figure}
\begin{center}\includegraphics[width=5in]{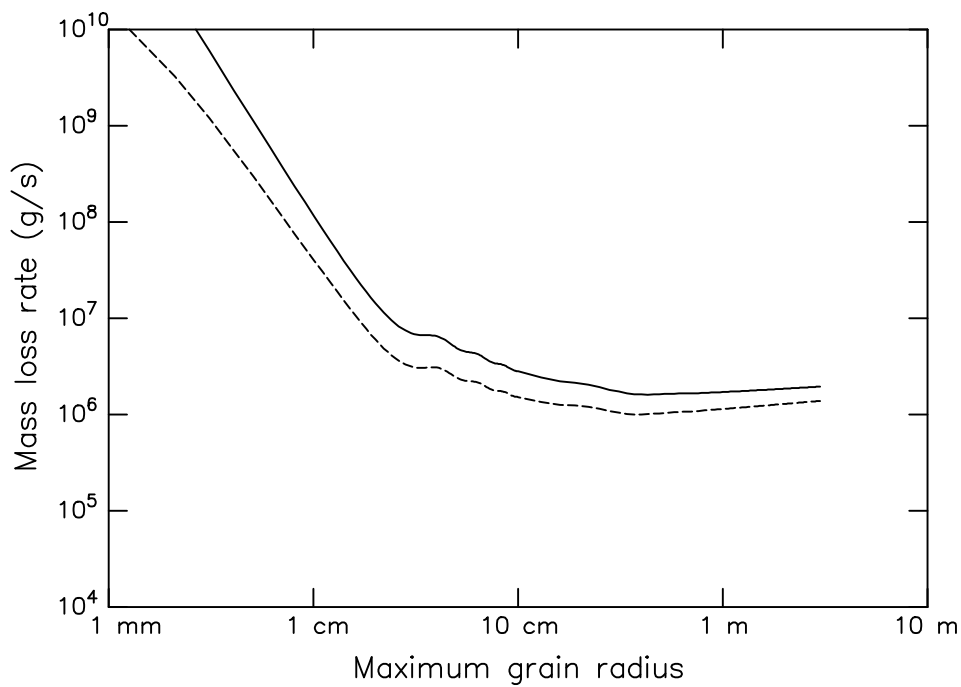}\end{center}
\caption{\label{fig:massloss}Mass-loss rate $\dot{M}$ (solid curve)
  computed from Eq.~4 assuming $C_{\rm v} = 36\,\rm cm^{1/2}$ \ms.,
  $\rho_{\rm g} = 0.5\,\rm g\,cm^{-3}$, $\alpha$ = 3.9, and $a_{\rm o}$
  = 1 $\mu$m.  Also shown (dashed) is the curve computed assuming these same
  parameters but with $a_{\rm o}$ = 1 mm.}
\end{figure}

Our basic model of stable and continuous grain ejection can also be
used to estimate the rate of mass loss $\dot{M}$ from the central
(nuclear) source. This involves calculation of the grain production
rate required to replenish the population of grains traversing and
ultimately exiting the radar beam. One approach is to calculate
$\dot{M}$ as a function of the gravity-limited size $a_{\rm m}$
assuming $C_{\rm v}$ to be an implicit function of $a_{\rm m}$ via
Eq.~\ref{eq:Cv}, as was done for IRAS-Araki-Alcock (Harmon et al.,
1989).  Alternatively, one can estimate $C_{\rm v}$ from the spectrum
width and then estimate $\dot{M}$ as a function of some
non-gravitational size cutoff $a_{\rm max}$ ($< a_{\rm m}$), as done
for Hyakutake (Harmon et al., 1997, 1999). For C/2001 A2 we take the
latter approach, given that our estimated gravitational cutoff $a_{\rm
  m}$ from Eq.~3 is much larger than $\lambda/2\pi$ and therefore
uninteresting as a constraint on $\dot{M}$.  In this case, assuming
isotropic grain ejection, the mass-loss rate required to give an
observed radar cross section $\sigma$ is given by
\begin{equation}
   \dot{M} = \sigma \left(\frac{8C_{\rm v} \rho_{\rm g}}{3}\right)\left[
          1-\left(\frac{a_{\rm o}}{a_{\rm max}}\right)^{4-\alpha}\right]
          {a_{{\rm max}}^{4-\alpha}\over(4-\alpha )\pi h I}
          \;\;\;\; ,
\end{equation}
where
\begin{equation}
   I = \int_{a_{\rm o}}^{a_{\rm max}}\frac{a^{(5/2-\alpha)}Q_{\rm b}(a)}
       {(1-a/a_{\rm m})^{1/2}}\; da \;\;\;\; .
\end{equation}
Here $h$ is the half-width of the radar beam at the comet and $Q_{\rm
  b}$ is the grain backscatter efficiency computed from Mie theory
assuming spherical grains.
In Fig.~\ref{fig:massloss} we show the calculated $\dot{M}$ curve for
$\alpha = 3.9$ and $a_{\rm o} = 1 \mu$m. We also plot a curve for
$a_{\rm o} = 1$ mm in order to show the $\dot{M}$ contribution from
the largest grains. The steep slope at low $a_{\rm max}$ reflects the
strong Rayleigh dependence for $Q_{\rm b}$. However, since we know
that $a_{\rm max}> 2$ cm from the depolarization, then $\dot{M}$ must
actually lie somewhere on the flat part of the curve. This means that
the large-grain $\dot{M}$ was $\sim$ 1--3$\times 10^6\,\rm g\,
s^{-1}$. For comparison, mass-loss rates in (1--10$\,\mu$m) dust from
this comet were estimated from optical spectra to be $\sim 3\times
10^5\,\rm g\,s^{-1}$ between late June and mid-July, 2001 (Schleicher
and Greer, 2001; Rosenbush et al., 2002), while gas production rates
during this same period varied between 1--2$\times 10^6\,\rm
g\,s^{-1}$ during quiescent conditions (Schleicher and Greer, 2001;
Lecacheux et al., 2001) to 6--9$\times 10^6\,\rm g\,s^{-1}$ during
outburst (Feldman et al., 2002). Thus, we find that the mass in large
grains is about 10 times the mass in 1--10 $\mu$m dust, and
comparable with the quiescent gas production. Table~\ref{tab:comets} shows that all of the comets with observable
grain coma show $\dot{M} \sim 10^6\,\rm g\,s^{-1}$. It is important to note
that our estimated mass-loss rates assume that the grains remain
intact from the time they leave the nucleus to the time they exit the
radar beam. If the grains were, in fact, disintegrating or evaporating
during this time, then the beam would not be filled and our $\dot{M}$
curves could underestimate the true mass loss and large grain
production. Conversely, if there is a continuous resupply of grains
from the breakup of larger boulders or nucleus fragments, then the
calculated $\dot M$ is an overestimate. The beam diameter was about
16000 km for these observations (see
Table~\ref{tab:radar}). For comparison, the IRAS-Araki-Alcock
radar echo was apparently contained within 2000 km
\citep{Campbell83:800,Harmon89:1071}, but as that comet showed no depolarized echo,
it presumably had smaller grains that could evaporate more quickly.

\section{Fragmentation and Grain Production}
The comet was observed to fragment several times before and after we observed it \citep{Sekanina02:679}. Although our assumption so far has been that the large grains were gas-drag ejecta from the surface of nucleus B, the fact that C/2001 A2 was a fragmenting object raises other possibilities. An interesting alternative is that the large grains were produced by, or were secondary products of, one or more of the observed episodes of nucleus fragmentation or splitting. Six fragments (A, G, C, D, E, F) were observed to separate from the main object (B) between late March and mid-June, 2001, and three visual-magnitude outbursts (presumed to be associated with some of the fragmentation events) were seen on March 28, May 10, and June 5 (Sekanina et al., 2002). A fourth outburst with no identifiable fragment association was observed on July 11, two days after the last radar observation. The possibility that the radar-reflecting grains were related in some way to the last (D-E-F) fragmentations of June 7--11 is consistent with the antisolar motion of these fragments. In fact, the observed 210--240$^{\circ}$ position angles observed for these separating fragments (relative to the main nucleus B) between mid and late June (Sekanina et al., 2002) also fit the radar-observed velocity, projected from its most antisolar possible direction, 165$^{\circ}$ from the Sun. Grains accelerated to the observed velocities during these fragmentations would have cleared out of the antenna beam by the time of the radar observations: Extrapolating the model of \nocite{Sekanina02:679}Sekanina et al. (2002) to our observation dates, fragment D should have left the beam, but fragments E and F may have remained within it. Fragments D-E-F had disappeared more than a week before these observations, but there may have been remnants. Moreover, grains and boulders produced in secondary breakups of fragments might tend to remain moving at the same slow ($\sim$\ms.) velocities as their parents. However, if grains produced in secondary fragmentation contained volatiles, then the rocket force produced by outgassing from their sunward sides could easily accelerate the grains down the antisolar tail to the observed velocities \nocite{Harris97:676,Hughes00:1,Bocklee01:1339}(Harris et al., 1997; Hughes, 2000; Bockel{\'e}e-Morvan et al., 2001). The grain velocity from the rocket force after a time $t$ is given by $V = 3 V_{\rm \! g} Z t / \pi a \rho_{\rm g}$.  Hence, grains can accelerate to the velocities shown in Fig.~\ref{fig:A2spec} in as little as an hour, depending on the grain volatile content, though it is not clear whether the acceleration is directional enough to produce the observed velocities without simply pinwheeling the grains. Even this pinwheeling could affect the resulting spectrum, by increasing the apparent velocity dispersion of the grains: a 10 cm radius grain rotating at 10 revolutions per second would give an echo bandwidth of up to 100 Hz. Pinwheeling would not explain the asymmetry in the spectrum, however. It is possible that a train of icy debris left over from the D-E-F fragmentation event, and undergoing continuous secondary fragmentation, could account for the radar-reflecting grain population observed long after the major fragmentation event itself. The gas produced in such secondary fragmentation could also account for the coma wings or ``arclets" that were observed symmetric about the tail axis of this comet. Jehin et al.\ (2002) reported arclets from observations on May 16 and July 13. In addition, Woodney et al.\ (2002) observed CN arcs on June 29-30 that were symmetric about a 250$^{\circ}$ position angle axis. Such arcs, which were also observed with comets Hyakutake and C/1999~S4 (LINEAR), may be produced when gas from a tailward debris train interacts with gas from the main nucleus (Harris et al., 1997; Rodionov et al., 1998; Boehnhardt, 2002). The possible analogy with Hyakutake is particularly relevant here, as Harmon et al.\ (1997) suggested that the radar coma echo from that comet may have come from the condensation of large grains invoked by Harris et al.\ (1997) to explain the gas arcs and other features.

It is possible that the radar coma grains were not products of the
D-E-F fragmentation but rather of some lesser fragmentation event that
occurred sometime between the D-E-F event and July 7 and which did not
produce any large visible fragments. For example, the CN arcs during
June 29-30 (Woodney et al., 2002) could be an indicator of a late-June
fragmentation event, although they may also have been a delayed
response to the D-E-F fragmentations. Another interesting possibility
is that the radar grains were produced in some unobserved
fragmentation event just preceding and triggering the July 11 outburst
(labeled ``outburst IV" by Sekanina et al., 2002). This seems
plausible given that outbursts I, II, and III (which Sekanina et al.\ 
attribute to secondary disintegration of fragmentation debris into
fine dust) all peaked a few days after their associated fragmentation
events. Hence, fresh fragmentation debris may have been present during
the radar observations and before the outburst itself. The arclet and
narrow tailward dust spike observed by Jehin et al.\ (2002) on July 13
might have been gas and dust products of the fragmentation of the same
large grains responsible for the radar coma echoes.  Since it would
have taken at least four days for a 1-cm grain to clear out of the
radar beam (at the maximum observed velocity of 50 \ms.), some of the radar coma grains associated with an
outburst-IV fragmentation could also have come directly from the
nucleus rather than from secondary fragmentation. The fact that the
days up to and including outburst IV were likely to have been a time
of chaotic activity suggests that one might have expected to see some
day-to-day changes in the radar echoes. This may, in fact, explain
some of the echo shape variations that were seen. Finally, it is worth
noting that the observation of the Hyakutake coma echo (Harmon et al.,
1997) was made only four days after a major outburst from that comet
(Schleicher and Osip, 2002). Hence, it is possible that the coma
echoes for both C/2001 A2 and Hyakutake reflected enhanced grain
injection associated with outbursts, although the time sequence for
the radar observations and the outbursts was reversed between the two
comets.

\section{Comparison with Other Comets}

Table~\ref{tab:OH} shows the model results from Section~3, along with total water production rates and velocities estimated using other observations.
The required gas flux per unit surface area $Z$ and particle acceleration parameter $C_{\rm v}$ are similar to that observed in Hyakutake, and quite different from that of Halley and IRAS-Araki-Alcock. For this purpose, C/2001~A2 and Hyakutake have high grain velocities (HGV) and Halley and IRAS-Araki-Alcock have low grain velocities (LGV). The model parameter $C_{\rm v}$ is about a factor of 6 higher and the parameter $Z$ about a factor of 100 higher in the HGV pair. We can also compare the total gas production from other observations. 

The observed gas velocities are all within a factor of 2 of 1 \kms., and do not correlate with the $C_{\rm v}$ parameter. Unfortunately, the estimate of gas velocity for IRAS-Araki-Alcock is very uncertain. The velocities should be considered further with data on more objects.

The model uses estimates of the gas flux per unit surface area $Z$. We compare these to the observed gas production rate $Q$ from passive observations of the OH radical at 1667~MHz. We use the estimated surface area $A$ of the comets and compute $Q/A$ for comparison. We expect $Q/A$ to be smaller than $Z$, because the comet is not active over its entire surface. We can then further compute an active fraction $f=(Q/A)/Z$. 

Table~\ref{tab:OH} shows that the HGV objects have areal gas fluxes $Q/A$ that are larger than the LGV objects. The difference is perhaps not conclusive for C/2001~A2, but recall that we only have an upper limit on the size of its nucleus. Since the area goes as $R^2$, the areal flux may easily be substantially larger.

The correlation between gas production (per unit area) and large-grain
velocity seems significant. They may both be related to the fact that
both C/2001 A2 and Hyakutake suffered breakup events \citep{Sekanina02:679,Schleicher03}, which likely produced a lot of gas and dust. C/2001 A2 and Hyakutake were also both much smaller than Halley and IRAS-Araki-Alcock, which may have influenced dust production as well.

\section{Conclusions}

The radar observations of C/2001 A2 (LINEAR) place an upper limit of
only 3 km for diameter of the nucleus and yet yield a grain-coma echo
with a cross section second only to that from Comet Halley. These
results, combined with other (non-radar) observations, establish
C/2001~A2 as a small but active comet similar to Hyakutake. Like
Hyakutake, this comet showed characteristic grain velocities of
several tens of meters per second, substantially higher than for the
grain-coma echoes from IRAS-Araki-Alcock and Halley.

Perhaps the most
important new result from these observations is the detection of
significant depolarization in the coma echo. The earlier radar
observations of comets Halley and Hyakutake showed hints of
depolarization, and mass-loss arguments had always implied the
dominance of cm-size or larger grains in the radar coma echoes.
Nevertheless, this solid detection of substantial echo depolarization
provides the first unambiguous radar evidence for coma grains larger
than the Rayleigh transition size $\lambda/2\pi$ = 2 cm. The marked
contrast with the IRAS-Araki-Alcock coma echo, which showed only 1.4\%
depolarization, is explainable with the simple gas-drag theory.  The
higher gas flux (inferred from the high grain velocities) and smaller
nucleus size for C/2001~A2 would be expected to give a gravitational
grain size limit much larger than the Rayleigh transition size. The
fact that C/2001~A2 was a fragmenting and outbursting object suggests
that a simple gas-drag model with a nucleus-centered source may not
tell the whole story for this comet. The observation of antisolar
fragmentation, gas arcs, and dust trains suggest that secondary grain
fragmentation and tailward self-acceleration could have been important
processes. These processes are at least consistent with the negative
Doppler offset of the echo and could also explain the high grain
velocities. We note that these same comments could apply just as well
to Comet Hyakutake, another fragmenting and outbursting comet showing
similar radar characteristics. Thus, while it is not clear that the
simple gas-drag model adequately describes C/2001~A2, it provides a
framework in which we can determine that the radar data require a
reasonably compact fan of ejected grains in a direction no closer than
45$^\circ$ to sunward, and also a steep (radius exponent $\alpha \sim
3.9$) size distribution of particles at cm-scales. Significant mass
loss for this comet occurs in the form of cm-size grains, suggesting
that large grains may also be present in other small, active comets.
Future comet radar observations and comparisons with
optical and other measurements should shed more light on
large-grain production processes in active comets.

\nocite{Biver1999:1850}
\nocite{Biver2002:5}
\nocite{Boehnhardt02}
\nocite{Campbell83:800}
\nocite{Eberhardt1999:45,Gunnarson2003:353}
\nocite{Feldman02:L91}
\nocite{Fulle95:622}
\nocite{Fulle97:1197}
\nocite{Fulle04}
\nocite{Harmon99}
\nocite{Jehin02:147}
\nocite{Jewitt97:1145}
\nocite{Lecacheux01:2}
\nocite{McDonnell91:1043}
\nocite{Rodionov98:232}
\nocite{Rosenbush02:423}
\nocite{Schleicher01:1}
\nocite{Schleicher02:210}
\nocite{Sykes86:1115}
\nocite{Whipple51:464}
\nocite{Woodney02:868}

\bibliographystyle{test}
\bibliography{short,mrabbrev,A2B}

\begin{deluxetable}{cccccccccccccc}
\rotate
\tablewidth{0pt}
%\tablecolumns{10}
\tabletypesize{\scriptsize}
\tablecaption{\label{tab:radar}Observing circumstances and radar measurements.}
\tablehead{UTC Mid-Time&RTT&Runs&$\Delta$&R&Beam&Angle&Integration&$\sigma_{\rm oc}$&$\sigma_{\rm sc}$&Total $\sigma$&$\mu_{\rm c}$&\\
&(s)&&(AU)&(AU)&(km)& &(s)&(km$^2$)&(km$^2$)&(km$^2$)}
\startdata
2001 July 7 08:53:44&262&13&0.262&1.132&16155&58$^\circ$&3210&4.10&1.20&5.30&0.27\\
2001 July 8 09:29:53&268&4&0.268&1.145&16497&55$^\circ$&\phantom{0}920&3.36&0.82&4.18&0.24\\
2001 July 9 08:12:14&274&8&0.274&1.158&16878&53$^\circ$&1920&4.98&1.47&6.45&0.30\\
Sum&&&&&&&6050&$4.43\pm 1.33$&$1.13\pm 0.34$&$5.56\pm1.67$&$0.28\pm 0.03$\\
\enddata
\tablecomments{Echo power results are given for each day
  of observations and the variance-weighted sum of all days. RTT is the round-trip light time to the
  object. ``Runs" is the number of transmit-receive cycles.
  Each cycle is 2 RTTs long. $\Delta$ is the Earth-comet
  distance. $R$ is the comet-Sun distance. ``Beam" is the diameter to
  the $1/\sqrt{2}$-power points of the telescope beam at the distance of
  the comet, or half-power considering the two-way path. ``Angle" is
  the Sun-comet-Earth angle. $\sigma_{\rm oc}$ and $\sigma_{\rm sc}$
  are the OC and SC
  radar cross-sections. $\mu_{\rm c}$ is the circular polarization
  ratio $\sigma_{\rm sc} / \sigma_{\rm oc}$. The random errors (due to
  thermal noise in the telescope and detector) for the cross-sections
  are a few percent. However, we estimate the systematic calibration
  uncertainties to be 30\%, as reported for the summed results. The uncertainties in the individual values are discussed in the text. The
  day-to-day variations in $\sigma_{\rm{oc}}$ are $> 20$\%, and probably reflect real changes in the comet.}
\end{deluxetable}

\begin{deluxetable}{lcccccccccc}%c}
\rotate
\tablewidth{0pt}
%\tablecolumns{10}
\tabletypesize{\scriptsize}
\tablecaption{\label{tab:comets}Parameters for all comets for which grain coma was detectable by radar.}
\tablehead{Comet&Date&$\Delta$&$\sigma_{\rm oc}^{\rm coma}$&$\mu_{\rm
    c}$&$2R$&$\alpha$&$a_{\rm m}$&$C_{\rm
    v}$&$Z$&$\dot{M}$\\%&$V_{\rm H_2O}$\\
&&(AU)&(km$^2$)&&(km)&&&($\rm cm^{1/2}\,$\ms.)&($\rm
g\,cm^{-2}\,s^{-1}$)&($\rm g\,s^{-1}$)}%&(\kms.)}
\startdata
C/2001 A2&2001-07-08&0.27&4.43&$0.28\pm 0.03$&2&3.9&$10\,$m&36&$70\times10^{-5}$&1--3$\times 10^6$\\%&2.19\tablenotemark{a}\\
Halley\tablenotemark{a}&1985-11-29&0.63&32&$0.52\pm 0.26$&10&3.5&$\ge2\,$cm&5\tablenotemark{c}&$0.8\times 10^{-5}\tablenotemark{b}$&0.5--$2\times 10^6$\\%&2.63\tablenotemark{d}\\
IRAS-Araki-Alcock\tablenotemark{c}&1983-05-11&0.03&0.80&$0.014\pm 0.003$&8.8&3.5&3$\,$cm&8&$1.2\times10^{-5}$&0.3--$1\times10^6$\\
Hyakutake\tablenotemark{d}&1996-03-24&0.11&1.33&$0.31\pm 0.12$&2.5&3.5&$\ge 1\,$cm&40&$40\times10^{-5}$&$1\times 10^6$\\%&2.6\tablenotemark{g}
C/2002 O6\tablenotemark{e}&2002-08-%
09&&1.1&$0.32\pm0.08$&\nodata&\nodata&\nodata&\nodata&\nodata&\nodata
\enddata
\tablecomments{The dates are approximate mid-times. $\Delta$ is the Earth-comet distance at the approximate mid-times. The coma cross sections all have systematic calibration uncertainties of
  30\%. 2$R$ is the comet diameter assumed in the modeling. $-\alpha$
  is the power-law exponent for the grain size distribution. $C_{\rm v}$ is the velocity scale factor, either from a
  model fit or from Eqs. 3 and 4. $Z$ is the gas flux at the
  surface required to lift particles of size $a_{\rm max}$ off the
  surface. $\dot{M}$ is the estimated mass in grains required to match
  the observed radar cross section using the model size
  distribution. \citet{Campbell89:1094} suggested
  $a_{\rm m}$ for Halley may actually be larger than
  2$\,$cm, giving a proportionately larger $Z$. The same argument
  applies to Hyakutake.}
%\tablenotetext{a}{\citet{Lovell}. Data from July 5.4.}
\tablenotetext{a}{\citet{Campbell89:1094}. The low SNR prevented a
reliable SC detection.}
\tablenotetext{b}{$C_{\rm v}$
  and $Z$ for Halley are computed from $a_{\rm max}$ and $2R$ using
  Eq. 3 and 4, $T= 200\,$K, $\rho_{\rm n}=1$, $\rho_{\rm
    g}=0.5$, and $C_{\rm D}=2$.}
%\tablenotetext{d}{\citet{Crovisier02}. Average value for the days of
%  radar observations.}
 \tablenotetext{c}{\citet{Harmon99}}
\tablenotetext{d}{\citet{Harmon97:1921}}
\tablenotetext{e}{\citet{Harmon04}. No modeling has been done for
  C/2002~O6. Its spectrum is qualitatively similar to the spectrum of
  C/2001~A2, with a similar bandwidth, but the SNR is lower.}
%\tablenotetext{g}{\citet{Crovisier02}. The value for March 25, because
%  the March 24 value was an outlier ($4.33\pm 0.78$).}
\end{deluxetable}

\begin{deluxetable}{cccccccc}
\tabletypesize\scriptsize
\tablecolumns{7}
\tablecaption{\label{tab:OH}Model and measured gas parameters.}
\tablehead{Comet&$r_{\sun}$&$C_{\rm v}$&$Z$&$Q$&$Q/A$&f&$V_{\rm H_2O}$\\
&(AU)&(cm$^{1/2}\rm \,m\,s^{-1}$)&$10^{18}$(cm${^{-2}}\,$s$^{-1}$)&$10^{28}\,$(s$^{-1}$)&$10^{18}$(cm${^{-2}}\,$s$^{-1}$)&&$\rm
km\,s^{-1}$}
\startdata
C/2001 A2&1.14&36&$23.4$&0.9&0.03\tablenotemark{a}&0.005\tablenotemark{a}&0.6\tablenotemark{b}\\
Halley&1.50&5&$0.268$&4--5\tablenotemark{c}&0.016&0.060&1.2\\
IRAS-Araki-Alcock&1.004&8&$0.402$&2.5&0.01&0.025&1-2.5\tablenotemark{d}\\
Hyakutake&1.08&40&$13.4$&30\tablenotemark{e}&1.52&0.113&1\tablenotemark{f}\\
%C/2002 O6&0.864\\
\enddata
\tablenotetext{a}{Since we only have an upper limit to the diameter, this is a lower limit to the areal flux and active fraction. If the diameter is really 1 km, then this flux is 0.30 and the active fraction is 0.05.}
\tablenotetext{b}{\citet{Lovell05}.}
\tablenotetext{c}{\citet{Gerard87}.}
\tablenotetext{d}{\citet{Irvine85}. The reported detections were very low SNR.}
\tablenotetext{e}{\citet{Gerard98}. The observations were from 1996 March 24.}
\tablenotetext{f}{\citet{Lovellthesis}.}
\tablecomments{The gas flux $Z$ is from Table 3, converted to molecules per second of H$_2$O. The total gas production rates $Q$ are measured values from the literature as indicated. $Q/A$ is converted to an areal flux using the size estimate in Table 2, assuming all gas is from the nucleus. The active fraction $f$ is estimated as ($Q/A)/Z$.}
\end{deluxetable}

%\begin{figure}
%\includegraphics{chisq.ps}
%\caption{\label{fig:chisq}Shade plot of $\chi^2$ for spectrum model fits. 
%  (0$^\circ$,0$^\circ$) represents the sunward direction. Darker
%  shading corresponds to higher $\chi^2$, so the locus of fan
%  direction giving good fits corresponds to the light band. The
%  dynamic range of $\chi^2$ is a factor of 20. For each grid point, a
%  model spectrum shape was computed. Then this model spectrum was
%  least-squares fit to the data with spectrum amplitude as the single
%  fit parameter. The $\chi^2$ value thus represents the fit residuals.
%  The fit assumes $\alpha = 3.9$, a $120^\circ$ fan width, and $C_{\rm
%    v} = 36\,\rm cm^{1/2}\,m\,s^{-1}$.}
%\end{figure}
\end{document}